\documentstyle[aps,prd,twocolumn,epsf,epsfig]{revtex}

\newcommand \bea{\begin{eqnarray}}
\newcommand \eea{\end{eqnarray}}
\newcommand \ga{\raisebox{-.5ex}{$\stackrel{>}{\sim}$}}
\newcommand \la{\raisebox{-.5ex}{$\stackrel{<}{\sim}$}}

\begin{document}
\twocolumn[\hsize\textwidth\columnwidth\hsize
\csname@twocolumnfalse%
\endcsname
\draft
\title{Second sound in Fermi gases at the BCS-BEC crossover}
\author{H. Heiselberg}
\address{Danish Defense Research Establishment, Ryvangsalle' 1, 
DK-2100 Copenhagen \O, Denmark}
\maketitle
\begin{abstract}
The thermodynamic potential is calculated for a uniform superfluid gas
of fermi atoms from the mean field BCS
equations including corrections from induced interactions,
Hartree-Fock energies and quasiparticle selfenergies.  The entropy,
specific heat and sound modes are calculated as function of temperature,
density and interaction strength from the BCS to the unitarity limit 
and around the BCS-BEC crossover. The
second sound speed is of particular interest as it is a clear signal
of a superfluid component and it determines the critical temperature.
\\ 
\end{abstract}
\vskip1pc]

Recent experiments probe systems of fermions near Feshbach resonances
by expansion\cite{Thomas,Bourdel,Regal,Ketterle,Jochim}, collective
modes \cite{Kinast,Bartenstein} and RF spectroscopy
\cite{Chin}. 
Interesting new strongly interacting or dense phases of
fermions and bosons are created, e.g., that associated with the
crossover from a superfluid or normal Fermi gas to a molecular BEC.
The experiments provide strong evidence for a superfluid state at low
temperatures. The next generation of experiments will measure
sound velocities in these systems. Second sound is particularly interesting
as it is a clear signal of a superfluid component and it also determines
the critical temperature.

Ho \cite{Ho} has studied the thermodynamics in the unitarity limit
at the Feshbach resonance and expressed a number
of thermodynamic quantities in terms of unknown universal parameters.
The purpose of this work is to extend the analysis to
all temperatures, densities and interaction
strengths in the BCS limit and around the unitarity limit, and to calculate
the thermodynamic quantities in terms the binding energy per particle
$E/N$ and the pairing gap $\Delta$ at zero temperature. These
quantities are known rather accurately as functions of density and
interaction strength from Monte Carlo \cite{Carlson,Casulleras} and
experimental results are compatible.  These functions will be treated
as input into a crossover model that is based on the mean field BCS
equations of Refs. \cite{Eagles,Leggett,NSR,Randeria}, however,
including important corrections in the crossover model from induced
interactions, self-energies and Hartree-Fock energies.  We will then
calculate the thermodynamic potential in the superfluid phase from
which the entropy, specific heat, first and in particular second sound
speeds can be calculated at low temperatures in the superfluid and
normal phases from the dilute BCS limit up to and around the unitarity
limit.

The mean field BCS equations of Refs. \cite{Eagles,Leggett,NSR,Randeria} have
become a standard reference for at least a qualitative
describing the BCS-BEC crossover as
function of density, interaction strength and temperature.
It describes a system of Fermi atoms of mass $m$ with
two spin states in spin equilibrium $n_\downarrow=n_\uparrow=n/2$ interacting
through a s-wave scattering length $a$.
The two-body interaction range is assumed to be short as compared to $|a|$
leaving only two length scales $a$ and the interparticle spacing 
(or $k_F^{-1}$). At zero temperature all 
the physics depend only on one variable which is
conveniently chosen as $x=1/(ak_F)$. 
It varies from $-\infty$ in the dilute BCS limit
through the unitarity limit at Feshbach resonance at
$x=0$ to $x\to +\infty$ in the molecular BEC limit.

The mean field BCS gap equation
\bea \label{gap}
  \frac{1}{g} = \sum_{\bf k} \left[
     \frac{1}{2E_k}-\frac{m}{\hbar^2k^2} -\frac{f_k}{E_k} \right] \,.
\eea
is valid for any coupling strength $g=-4\pi \hbar^2a/m$
and can thus at least qualitatively
describe the smooth crossover from a BCS state to a molecular BEC.
As usual $\varepsilon_k=\hbar^2k^2/2m-\mu_\Delta $ and
$E_k=\sqrt{\varepsilon_k^2+\Delta^2}$. Note that the
chemical potential $\mu_\Delta $ does not include
Hartree-Fock energies and the quasiparticle energy does not contain
any self-energy.
The thermal distribution function is
$f_k=(\exp(E_k/T)+1)^{-1}$ in units where $k_B=1$.
With the equation for number density conservation
\bea \label{n}
  n = \sum_{\bf k}  \left[ 1-\frac{\varepsilon_k}{E_k} 
        +\frac{\varepsilon_k}{E_k} \frac{f_k}{2}   \right] \,,
\eea
the gap and chemical potential can be calculated as function of
density, temperature and interaction strength.  At zero temperature
the last term in Eqs. (\ref{gap}) and (\ref{n}) vanish and the gap and
chemical potential are easily calculated as function of 
$x=1/(ak_F)$ (see Refs. \cite{Leggett,Randeria} and Fig. 1).

In the dilute BCS limit $\Delta\ll\mu_\Delta \simeq E_F$, 
where the Fermi energy is
$E_F=\hbar^2k_F^2/2m$, and the Fermi momentum $\hbar k_F$ is given in
terms of the density $n=k_F^3/3\pi^2$.
The sum in the gap equation 
is at zero temperature
simply $(mk_F/2\pi^2)\ln(\kappa E_F/\Delta_0)$, where $\kappa=8/e^2$.
The superfluid gap at zero temperature $\Delta_0=\Delta(T=0)$ becomes
\bea \label{Gorkov}
  \Delta_0=\kappa E_F \exp\left(\frac{\pi}{2ak_F}\right) \,.
\eea
Gorkov, however, found that induced interactions lead to a higher order 
correction such that: $a^{-1}\to a^{-1}-2k_F\ln(4e)/3\pi$  \cite{Gorkov}.
In the above gap equation, where the l.h.s. is proportional to
$x=1/(ak_F)$, this correction corresponds to adding or shifting
$x$ by the amount $2\ln(4e)/3\pi$.
As result the  Gorkov gap has $\kappa=(2/3)^{7/3}$ in Eq. (\ref{Gorkov}). 
In the following we shall add this constant 
shift correction not only in the dilute BCS limit but generally at all $x$.
Therefore the resulting gap and chemical
potential as shown in Fig. 1
both deviate from standard results by the above shift in  $x$.
The resulting gap is exact in the dilute BCS limit and generally 
in good agreement with Monte Carlo calculations
\cite{Carlson} up to $x\la 1$.

Neither self-energies nor Hartree-Fock energies 
are included in the mean field BCS equations.
Monte Carlo calculations provide detailed insight in their contributions
relative to the effect of pairing because
pairing can be included or excluded in the trial wave functions.
The energy per particle
$E/N=(3/5)E_F(1+\beta)$ is expressed in terms of the universal function
$\beta(x)=E_{int}/E_{kin}$ \cite{HH}. In the
unitarity limit $\beta(x=0)=-0.46$ without pairing but $\beta(x=0)=-0.56$ with
pairing in the Monte Carlo calculations of Ref. 
\cite{Carlson} at zero temperature. 
Therefore, the dominant contribution to the binding energy
and chemical potential does not come from pairing
but from other correlations in the wave function
that contribute to the Hartree-Fock energy. 
We shall therefore distinguish between 
$\mu_\Delta $, which includes only the pairing effects, and the correct
chemical potential 
\bea
  \mu=\left(\frac{\partial E}{\partial N}\right)_{V,S}
    =E_F\left(1+\beta-\frac{1}{5}x\frac{d\beta}{dx} \right) \,,
\eea
which includes both pairing and Hartree-Fock energies. Both are shown in
Fig. 1. On the BCS side and even in the unitarity limit $\mu_\Delta $ 
differs from $E_F$ by a small amount only. This is compatible with
the small deviation between the chemical potentials calculated by
Monte Carlo with and without pairing. On the BEC side $x=1/(ak_F)>0$
the pairing contribution to binding energies become increasingly important,
and the chemical potential rapidly drops toward the molecular binding
energy $\mu_\Delta \simeq\mu\simeq -\hbar^2/2ma^2$ in the dilute BEC
limit.
Recent experiments on expansion energies 
\cite{Thomas,Bourdel,Regal,Ketterle,Jochim} and
collective modes in traps \cite{Kinast,Bartenstein} are compatible
with the Monte Carlo calculations.

In both the hydrodynamic limit and for a superfluid gas 
the first sound is given by the adiabatic sound speed
\bea
   u_1^2 = \frac{n}{m} \left(\frac{\partial \mu}{\partial n}\right)_{V,S}
      = \frac{1}{3}v_F^2 
  \left[ 1+\beta-\frac{3}{5}x\beta' +\frac{1}{10}x^2\beta''\right] \,,
\eea
where $v_F=\hbar k_F/m$, $\beta'=d\beta/dx$, etc.
In the dilute BCS limit and at low temperature 
$u_1=v_F\sqrt{(1+(2/\pi)ak_F)/3}$. 
In the unitarity limit
$u_1=\sqrt{2\mu(x=0)/3m}=v_F\sqrt{(1+\beta(0))/3}\simeq 0.37v_F$. 
In the dilute BEC limit $u_1=\sqrt{(\pi/2)\hbar^2na_m/m^2}$, 
where $a_m\simeq 0.6a$ is the molecular scattering length 
\cite{Petrov,Casulleras}.
The first sound speed from Monte Carlo calculations \cite{Casulleras}
is plotted in Fig. 1. 

In a normal Fermi liquid the first sound speed is expressed in terms of the
Landau parameters as: $u_1^2=(v_F^2/3)(1+F_0)/(1+F_1/3)$, at low
temperatures. The Fermi liquid theory can be generalized to finite
temperatures $T\la 0.5T_F$ and arbitrary relaxation times
\cite{Ravenhal}. The effective mass
$m^*=(1+F_1/3)m$ may be determined from, e.g., the specific heat as shown
below, and
$F_0$ can then be determined from $u_1$ or equivalently $\beta(x)$. 
In the dilute BCS limit $m^*/m=1+[8(7\ln2-1)/15\pi^2]a^2k_F^2$.
In the mean field BCS equation $m^*=m$ because selfenergies are not included. 
We shall in the following assume that selfenergies can be included in
the quasiparticle energies such that they lead to an effective mass at
the Fermi surface. Eventually $m^*$ will have to be measured and/or calculated
by Monte Carlo as in the cases of $\Delta_0$ and $\mu$.
The Landau parameter $F_0$ varies from zero  in the BCS limit to $-1$ in the
BEC limit where, however, the liquid will only be in the normal state for
temperatures above the critical temperature. For a BEC this condition is
$T\ge T_c=(n/2\zeta(3/2))^{2/3}\pi/m\simeq0.218E_F$.
Fermi liquid theory also describes collisions and the transition between the
hydrodynamic (first) sound and the collisionless (zero) sound as the 
collision rate decrease. The Landau damping can also be calculated and for
$-1\le F_0\le0$ the zero sound mode becomes purely imaginary in the
collisionless limit.

\vspace{-0.5cm}
\begin{figure}
\begin{center}
\psfig{file=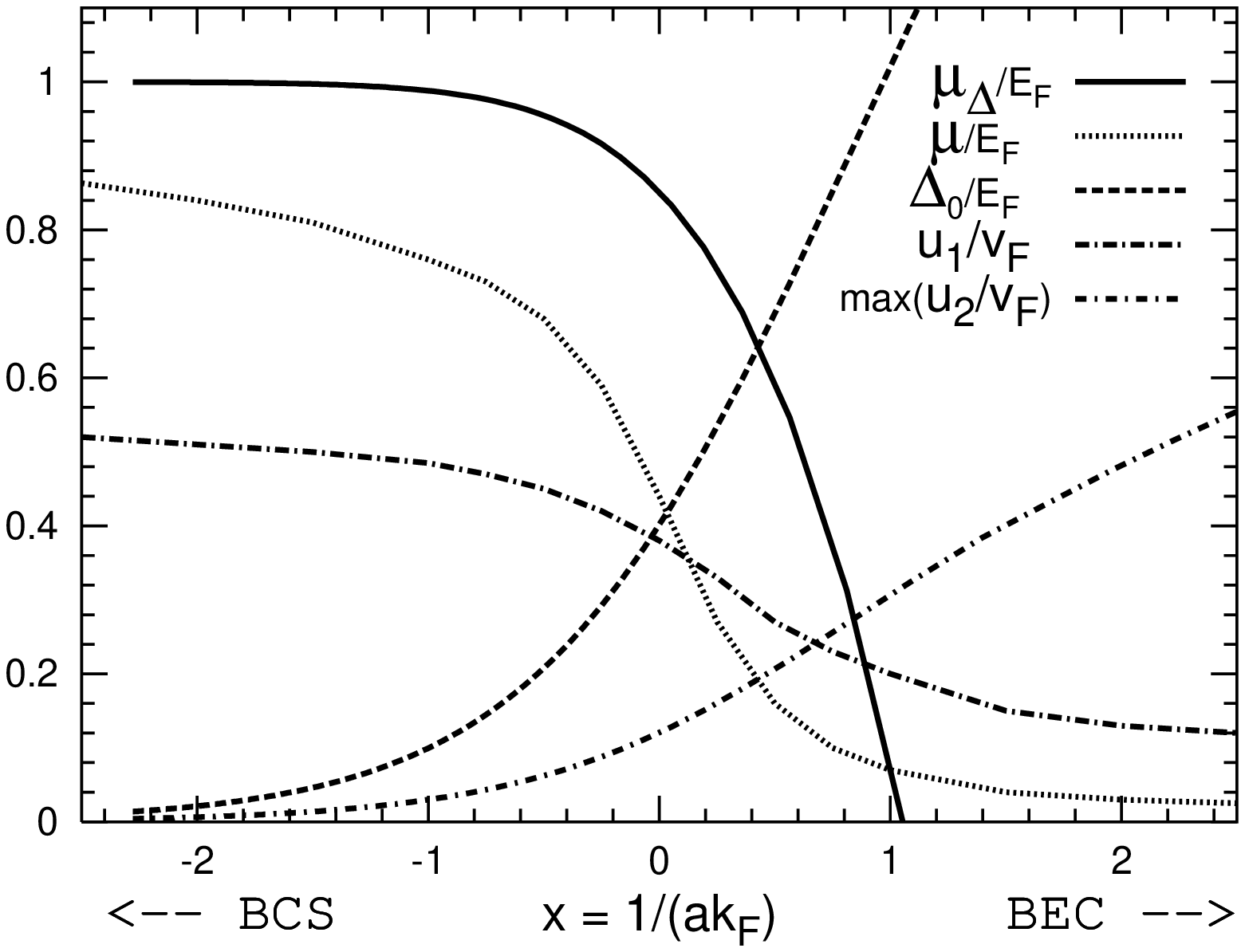,height=8.0cm,angle=-90}
\vspace{.2cm}
\begin{caption}
{The chemical potential $\mu_\Delta $ and pairing gap $\Delta_0$ 
in the crossover model,
and $\mu$ from Monte Carlo calculations [11]
(all in units of $E_F$ and at $T=0$).
The first and second sound speed at maximum ($T\simeq 0.7T_c$ assuming
$m^*=m$) are also shown in units of $v_F$. 
}
\end{caption}
\end{center}
\label{f1}
\end{figure}
\vspace{-.2cm}

 In the following we shall refer to the mean field BCS equations with
the above described 
corrections from induced interactions, Hartree-Fock energies and
effective mass as simply the ``crossover model''. It
contains the correct pairing and mean field in the dilute BCS
limit to leading orders, and is also correct in the unitarity limit to
a good approximation. It does not solve the full many-body problem of
strongly interacting Fermi gases but should regarded as an approximate
model that includes some of the most important physics. As will shown
in the following it has the virtue that it produces definite predictions
for a number of observables which are about to be measured.

The gap at finite temperature is found by solving the gap equation.
It is a major simplification that we for most purposes treat the
pairing gap as being small as compared to the Fermi energy for
$x\la0$.  The next order correction to, e.g. the gap itself turns out
to be $(\Delta/2E_F)^2$, which even in the unitarity limit and at zero
temperature is less than 10\% in both the crossover model and Monte
Carlo calculations.  To the same order in $\Delta/E_F$ we can also
approximate $\mu_\Delta\simeq E_F$ in the gap equation.

Inserting the zero temperature gap of Eq. (\ref{Gorkov}) 
back into the gap equation (\ref{gap}), we arrive at 
the expression for the finite temperature gap
\bea \label{lnD}
  \ln\left(\frac{\Delta_0}{\Delta(T)}\right) = 
  2\int_0^\infty d\varepsilon_k \frac{f_k}{E_k} \,.
\eea
Because the gap is small, only the quasi particle
energies around the Fermi surface only
contribute to the integrals. Therefore, the effective mass in
the quasiparticle  selfenergy
near the Fermi surface appears in the level density.
The effective masses should also be included
selfconsistently in $\Delta_0$.
Therefore $\Delta_0$ may deviate from the Gorkov gap, and $\Delta_0$ should
be taken from experiment or Monte Carlo calculations as will be assumed in
the following.

The temperature dependence of the gap resulting from Eq.
(\ref{lnD}) is plotted in Fig. 2.
It vanishes above the critical temperature,
$T_c=(\gamma/\pi)\Delta_0$,
and has the form $\Delta=\pi T_c\sqrt{8/7\zeta(3)}\sqrt{1-T/T_c}$ for
$T_c-T\ll T_c$. Near zero temperature 
$\Delta=\Delta_0-\sqrt{2\pi\Delta_0T}\exp(-\Delta_0/T)$.

The thermodynamic functions can be calculated from the
thermodynamic potential per volume $\Omega=-P$. We 
make the standard assumption that the Hartree-Fock terms in the superfluid
$\Omega_s$ and normal state $\Omega_n$ thermodynamic potentials
are the same.
The difference is then given in terms of the pairing coupling as
\bea\label{OS}
  \Omega_s &=& \Omega_n + \int_0^\Delta d\Delta' \Delta'^2 
         \frac{d(1/g)}{d\Delta'}  \,.
\eea
The first order temperature correction to the thermodynamic potential
in the normal phase is $\Omega_n=-N(0)\pi^2 T^2/3$, where
$N(0)=m^*k_F/2\pi^2$ is the level density.  It should be noted that
the above thermodynamic potential is assumed to be valid also in
unitarity regime, where the Hartree-Fock terms are of order
$\beta E_F$. However, as
discussed above the pairing has relatively small effect on the
chemical potentials and we may therefore expect that
the difference between Hartree-Fock
terms in the superfluid and normal phases also remains small.

Reinserting the gap of Eq. (\ref{lnD}) 
in the gap equation we find for the coupling 
\bea \label{1g}
  \frac{1}{g} =  -N(0) \left[\ln\left(\frac{\Delta}{\kappa \mu_\Delta} \right) 
                +2\int_0^\infty d\varepsilon_k \frac{f_k}{E_k} \right] \,.
\eea
Note that $\kappa$ now included induced interactions and possible
effective mass corrections such that the correct $\Delta_0$ of
Eq. (\ref{Gorkov}) is reproduced.
Inserting the coupling of Eq. (\ref{1g}) into the thermodynamic potential
(\ref{OS}) and again exploiting that $\Delta\ll\mu_\Delta $,
it reduces to
\bea \label{OT}
   \Omega_s &=& -N(0)\left[ \Delta^2\left(\frac{1}{2}+
     \ln\frac{\Delta_0}{\Delta}\right)
       -4T\int_0^\infty d\varepsilon_k \ln(1-f_k) \right] \nonumber\\
   && 
\eea
The crossover model thus arrives at the standard expression for $\Omega_s$
and therefore also the standard
entropy $S_s = -(\partial\Omega_s/\partial T)_{V,\mu}$,
and specific heat
$C_s = T(\partial S_s/\partial T)_{V,\mu}$ in superfluid phase.
In the normal phase $S_n=C_n=N(0)2\pi^2 T/3$.
At low temperatures 
$S_s=N(0)\sqrt{2\pi\Delta_0^3/T}\exp(-\Delta_0/T)$ and
$C_s=N(0)\sqrt{2\pi\Delta_0^5/T^3}\exp(-\Delta_0/T)$. Near $T_c$:
$S_s/S_n(T)=1-(1+\xi)(1-T/T_c)$  and
$C_s/C_n(T_c)=\xi -3.77(1-T/T_c)$. 
Here $\xi=1+12/7\zeta(3)\simeq2.43$ is the superfluid 
specific heat relative to the normal one at $T_c$.

Finally, we need the superfluid (London) density
\bea
   n_s &=& n\left(1+2\int^\infty_0 d\varepsilon_k \frac{df_k}{dE_k} \right)\,, 
\eea
where $n=n_s+n_n$ is the total density.
At low temperatures $n_s/n=1-\sqrt{2\pi\Delta_0/T}\exp(-\Delta_0/T)$
whereas  $n_s/n=2(1-T/T_c)$ near $T_c$.

\vspace{-0.5cm}
\begin{figure}
\begin{center}
\psfig{file=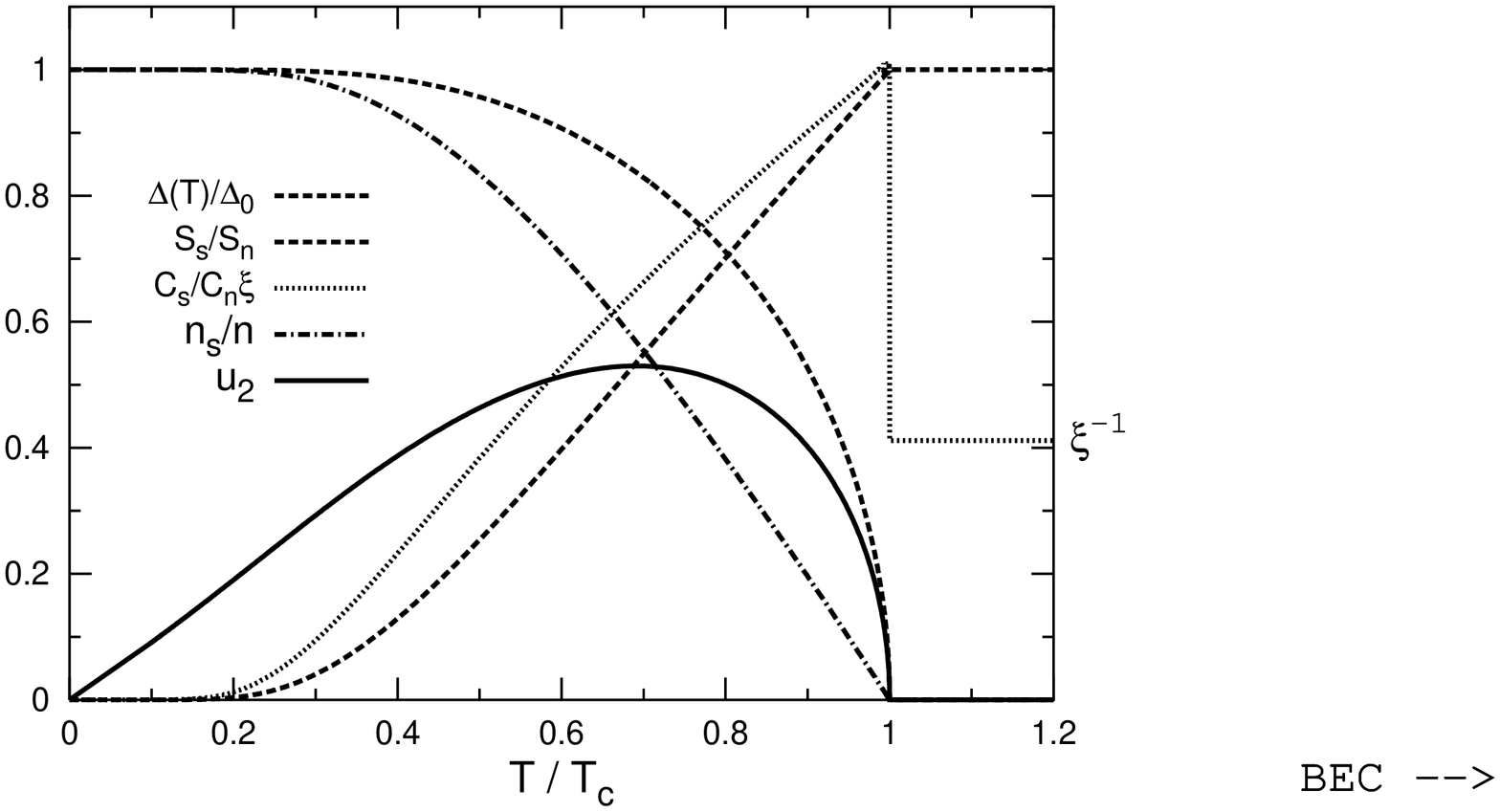,height=8.0cm,angle=-90}
\vspace{.2cm}
\begin{caption}
{Thermodynamic quantities for a superfluid as calculated in the crossover 
model vs. $T/T_c$.
Shown are the gap $\Delta(T)/\Delta_0$, entropy $S_s(T)/S_n(T)$, 
specific heat $C_s(T)/C_n(T)\xi$, and superfluid density $n_s(T)/n$.
The second sound is plotted in units of $v_F\sqrt{m^*/m}T_c/E_F$.
}
\end{caption}
\end{center}
\label{f2}
\end{figure}
\vspace{-.2cm}

We now have all the thermodynamic quantities available for calculating the
second sound speed
\bea
  u_2^2 = \frac{n_s}{n_n}\frac{S_s^2T}{mnC_s} \,.
\eea
as shown in Fig. 2.
At low temperatures the second sound is linear in temperature
\bea
  u_2=\frac{\sqrt{3}}{2} \sqrt{\frac{m^*}{m}}\frac{T}{E_F}v_F
    \,,\quad\quad T\ll T_c \,.
\eea
Around $T\simeq 0.7T_c$
the second sound speed has a broad maximum of
\bea
u_2\simeq 0.53 v_F \sqrt{\frac{m^*}{m}}\frac{T}{E_F} \,,\,\quad T\sim 0.7T_c\,.
\eea 
Near the critical temperature $T_c-T\ll T_c$ second sound
\bea
   u_2=\frac{\pi}{\sqrt{\xi}} \sqrt{\frac{m^*}{m}}\frac{T}{E_F}
       \sqrt{1-\frac{T}{T_c}} v_F   \,,
\eea
has the characteristic $u_2\propto \sqrt{1-T/T_c}$ behavior
and the vanishing point determines $T_c$. 

Ho \cite{Ho} predicted the scaling behavior of the above thermodynamic
quantities in the unitarity limit ($x=0$) near $T_c$ from universal
scaling laws. In his analysis 
the $u_2\propto\sqrt{1-T/T_c}$ dependence near $T_c$ and $x=0$ was found
but quantitatively it depended on unknown parameters.
In comparison the crossover model can be applied
to all temperatures, densities and interaction
strengths as long as $x\la 0$. The results are given in terms of the functions
$\Delta_0(x)$, $\mu(x)$ and $m^*(x)$, of which the first two
are known from Monte Carlo calculations and experiments as discussed above.

The crossover model breaks down in the BEC
limit, $x\ga 0.5$, because the gap becomes large, the chemical potential
becomes negative, gaussian fluctuations become important
at $T_c$ \cite{Randeria}, and 
pseudogaps appear \cite{Levin,Torma}. The second sound with maximum
$u_2\simeq 0.53 v_F (T_c/E_F)$ can
therefore not be trusted on the BEC side for $x\ga0.5$ in the present model.
If the first and second sound continue to decrease and increase
respectively as we go from the BCS to the BEC limit (see Fig. 1), i.e.
as $x\ga 0$, they would cross. However, 
the two sound modes are coupled and undergo avoided crossing.
This is known to occur in a BEC at the temperature 
$T^*=|g|n$ \cite{Zaremba,PS}. Below this temperature ($T\la T^*$)
the Bogoliubov condensate mode $u^2=|g|n_s(T)/m$ is first
sound whereas above it is second sound. Above the first sound speed
is the classical result $u_1=0.925\sqrt{T/m}$.
At the crossings, where the normal and superfluid components couple strongly,
we may expect stronger damping of both sound modes.

In summary, the mean field BCS equations were corrected with induced
interactions, Hartree-Fock energies and self-energies. The resulting
crossover model is approximately correct at all temperatures,
densities and interaction strengths up to $x\la0.5$, and exact in the
dilute BCS limit. A number of thermodynamic quantities were
calculated for gas of Fermi atoms with superfluid and normal
components.  Second sound is particularly important because it only
appears in a superfluid gas and it reveals $T_c$ where it vanishes.
The predicted sound speeds in the superfluid phase
can be tested in upcoming experiments. The validity of crossover model
will be put to a test in the unitarity limit and measurements of the
sound modes on the molecular BEC side will provide important information 
on new phenomena as pseudogaps, quasiparticle energies, etc.



\vspace{-0.4cm}
\begin{thebibliography}{99}

\bibitem{Thomas}
K. M. O'Hara, S. L. Hemmer, M. E. Gehm, S. R. Granade, J. E. Thomas,
Science {\bf 298} (2002) 2179.
M. E. Gehm et al.,
Phys. Rev. A {\bf 68}, 011401 (2003); cond-mat/0304633 .
\bibitem{Bourdel} T. Bourdel et al., Phys. Rev. Lett.  {\bf 91}, 020402 (2003);
cond-mat/0403091; J. Cubizolles et al., cond-mat/0308018

\bibitem{Regal}
C. A. Regal et al., Phys. Rev. Lett. {\bf 90}, 230404 (2003); cond-mat/0305028;
M. Greiner, C.A. Regal, D.S. Jin, Nature {\bf 426}, 537 (2003);
Phys. Rev. Lett. {\bf 92}, 040403 (2004).

\bibitem{Ketterle}  M.W. Zwierlein et al.,
Phys. Rev. Lett. {\bf 91}, 250401 (2003); ibid. {\bf 92}, 120403 (2004).
S. Gupta et al., Science {\bf 300}, 47 (2003).

\bibitem{Jochim} S. Jochim et al.,
 Phys. Rev. Lett.  {\bf 91}, 240402 (2003); M. Bartenstein et al., 
ibid.   {\bf 92}, 120401 (2004)

\bibitem{Kinast} J. Kinast et al.,  Phys. Rev. Lett.  {\bf 92}, 150402 (2004)
\bibitem{Bartenstein} M. Bartenstein et al., 
Phys. Rev. Lett.  {\bf 92}, 203201 (2004)

\bibitem{Chin} C. Chin, M. Bartenstein A. Altmeyer, S. Riedl, S. Jochim, 
J. Hecker Denschlag, R. Grimm, Science {\bf 305} 1128 (2004).


\bibitem{Ho} T-L. Ho,  Phys. Rev. Lett. {\bf 92}, 090402 (2004)


\bibitem{Carlson} J.~Carlson, S-Y. Chang, V.~R. Pandharipande, K.~E. Schmidt, 
Phys. Rev. Letts. {\bf 91}, 50401 (2003).   
S-Y. Chang et al., physics/0404115;  nucl-th/0401016.

\bibitem{Casulleras} G.E. Astrakharchik, J. Boronat, J. Casulleras, S. Giorgini, cond-mat/0406113.

\bibitem{Eagles} D.M. Eagles, Phys. Rev. {\bf 186}, 456 (1969).

\bibitem{Leggett} A.J. Leggett, in {\it Modern Trends in the Theory of Condensed Matter}, ed. A. Pekalski and R. Przystawa, Lect. Notes in Physics Vol. 115
(Springer-Verlag, 1980), p. 13.

\bibitem{NSR} P. Nozi\`eres and S.~Schmidt-Rink, J. Low
Temp. Phys. {\bf 59}, 195 (1982).  

\bibitem{Randeria}
C.\ A.\ R.\ S\'a de Melo, M.\ Randeria, and J.\ R.\ Engelbrecht,
Phys.\ Rev.\ Lett.\ {\bf 71}, 3202 (1993).
M. Randeria, in ``Bose-Einstein Condensation'', Ed. A. Griffin, D.W. Snoke, S. Stringari, Cambridge Univ. Press 1995, p. 355. 

\bibitem{Gorkov} L.~P.~Gorkov \& T.~K.~Melik-Barkhudarov, 
JETP {\bf 13}, 1018 (1961);
 H. Heiselberg, C.~J. Pethick, H. Smith and L. Viverit,
Phys. Rev. Letts.  {\bf 85}, 2418 (2000).

\bibitem{HH} H.~Heiselberg,  Phys. Rev. {\bf A 63}, 043606 (2001);
J. Phys. B: At.Mol.Opt.Phys. {\bf 37}, 1 (2004).

\bibitem{Petrov} D.S. Petrov, C. Salomon, and G.V. Shlyapnikov, 
cond-mat/0309010.

\bibitem{Ravenhal} H. Heiselberg, C.J. Pethick, D.G. Ravenhall,
             Annals of Physics {\bf 223}, 37 (1993).

\bibitem{Levin} J. Stajic, J. N. Milstein, Qijin Chen, M. L. Chiofalo, 
M. J. Holland, K. Levin, Phys. Rev. A {\bf 69}, 063610 (2004).

\bibitem{Torma} J. Kinnunen, M. Rodriguez, P. T\"orma, cond-mat/0405633.

\bibitem{Zaremba}A. Griffin and E. Zaremba, Phys. Rev. A {\bf 56}, 4839 (1997).

\bibitem{PS} C.J. Pethick and H. Smith, {\it Bose-Einstein Condensation
in Dilute Gases}, Cambridge Univ. Press, 2002.


\end{thebibliography}
\end{document}